# Optical wavelength conversion of high bandwidth phase-encoded signals in a high FOM 50cm CMOS compatible waveguide

Francesco Da Ros,[1, a)] Edson Porto da Silva,[1] Darko Zibar,[1] Sai T. Chu,[2] Brent E. Little,[3] Roberto Morandotti,[4, 5, 6] Michael Galili,[1] David J. Moss,[7] and Leif K. Oxenløwe[1]

[1] *DTU Fotonik, Technical University of Denmark, DTU, Kongens Lyngby, 2800, Denmark, DK*

[2] *Department of Physics and Materials Science, City University of Hong Kong*

[3] *Xi'an Institute of Optics and Precision Mechanics, CAS, Xi'an, China PRC*

[4] *INSR - Énergie,Matériaux et Télécommunications, 1650 Blvd Lionel Boulet, Varennes (Québec), J3X1S2, Canada*

[5] *Institute of Fundamental and Frontier Sciences, University of Electronic Science and Technology of China, Chengdu 610054, China*

[6] *National Research University of Information Technologies, Mechanics and Optics, St. Petersburg, Russia.*

[7] *Centre for Microphotonics, Swinburne University of Technology, Hawthorn, VIC 3122, Australia*

(Dated: 30 January 2017)

## Abstract

We demonstrate wavelength conversion of QAM signals including 32-GBd QPSK and 10-GBd 16-QAM in a 50cm long high index doped glass spiral waveguide. The quality of the generated idlers over a 10-nm bandwidth is sufficient to achieve a BER performance below the HD-FEC threshold ($< 3.8 \times 10^{-3}$), with an OSNR penalty of less than 0.3 dB compared to the original signal. Our results confirm that this is a promising platform for nonlinear optical signal processing – a result of both very low linear propagation loss ($< 0.07$ dB/cm) and the large material bandgap that ensures negligible nonlinear loss at telecom wavelengths.

## I. INTRODUCTION

All-optical wavelength conversion is one of the most promising signal processing techniques for future optical wavelength division multiplexed (WDM) networks, as it enables more effective and efficient wavelength management in the network, especially in terms of routing functionality. Currently, routing in optical networks relies on electronics, thus requiring optical to electrical (O/E) and electrical to optical (E/O) conversion. It has been shown that lower wavelength blocking probability can be achieved by performing routing directly in the optical domain[1]. From this perspective, wavelength conversion plays a key role as a collision avoidance method, while providing more advanced functionality such as path protection, dispersion compensation and Kerr nonlinearity mitigation through optical phase conjugation[2]. However, in order for wavelength converters to be applied to optical networks, modulation format independent operation needs to be ensured. Furthermore, as coherent transmission systems using quadrature phase-shift keying (QPSK) and 16-quadrature amplitude modulation (16-QAM) are already commercially deployed, the stringent requirements in terms of optical signal-to-noise ratio (OSNR) need to be addressed. Significant progress has been made for wavelength converters based on four-wave mixing (FWM) in highly nonlinear fibers (HNLFs), with conversion of dual-polarization 64-QAM signals recently being demonstrated[3]. However, nonlinear processing in HNLFs is limited by stimulated Brillouin scattering (SBS), and even though techniques such as fiber straining[4] or phase dithering in counter-phasing operations[3] have been shown to mitigate the impact of SBS, they come at the price of a significant increase in device complexity. A nonlinear platform with higher SBS immunity would therefore be highly advantageous in providing higher performing schemes. Integrated wavelength converters based on compact waveguide devices offer precisely this, with the added benefits of stability, reduced footprint, cost, and improved scalability.

Previous reports of integrated wavelength converters for advanced modulation formats have achieved wavelength conversion of QPSK[5-7], 16-QAM[8,9], and up to 128-QAM orthogonal frequency division multiplexing (OFDM)[10] using silicon[5,7-10] and silicon-germanium[6] waveguides. However, these materials are well known to be limited by nonlinear absorption at telecom wavelengths, thus motivating the search for other nonlinear material platforms such as AlGaAs[11], silicon nitride[12], amorphous silicon[13] and high index doped glass[14–17] that exhibit a much higher nonlinear figure-of-merit (FOM)[14] in the telecom wavelength range.

Here, we extend our previous system demonstrations of wavelength conversion[17,18] by providing a detailed static (i.e. under continuous wave (CW) operation), and dynamic investigation of wavelength conversion in a 50-cm long high index doped glass spiral waveguide focusing on the performance of the converter for QAM signals. The achievable conversion efficiency (CE) and bandwidth are thoroughly investigated, highlighting the impact of the pump wavelength and power level. The results of this investigation are then applied to demonstrate wavelength conversion of 32-GBd QPSK and 10-GBd 16-QAM over a full 10-nm idler bandwidth, with a maximum signal-idler separation of 20 nm. In particular, we show that only a moderate pump power of 22 dBm is required to achieve a bit error ratio (BER) performance below the hard decision forward error correction (HD-FEC) threshold (BER=$3.8\times10^{-3}$) after wavelength conversion in a 50-cm long spiral waveguide. We achieve an optical signal-to-noise ratio (OSNR) penalty, with respect to the back-to-back signal, of less than 0.3 dB over the entire 10-nm idler bandwidth, confirming the strong potential of high index doped glass waveguides for nonlinear optical signal processing. Comparison with a 3-cm long waveguide is further provided, showing a similarly low penalty but lower OSNR margin on the generated idler, thus demonstrating the advantage of the longer spiral structure.

## II. WAVEGUIDE PROPERTIES

A scanning electron microscope (SEM) image of the waveguide is shown in Fig. 1(a) prior to deposition of the silica over-cladding. The 50cm long spiral waveguide (core refractive index n$\approx$ 1.8) is made of a variation of silicon oxynitride (SiON) and has a cross section of 1.45$\mu$m$\times$1.50$\mu$m, surrounded by a silica cladding with an upper cladding thickness of 4 $\mu$m [14].

The final device was pigtailed with fibers glued directly to the sample using V-grooves for alignment. The total insertion loss, including propagation loss ( $\leq$ 0.07 dB/cm), mode converter loss ($\approx$ 1.2 dB/facet) and pigtail splicing loss is approximately 10.5 dB. The group velocity dispersion is estimated to be around + 8 ps/nm·km at 1550 nm for the TM mode, i.e. anomalous, as shown in Fig. 1(b), and the nonlinear parameter $\gamma$ is 220 W$^{-1}$km$^{-1}$ [14].

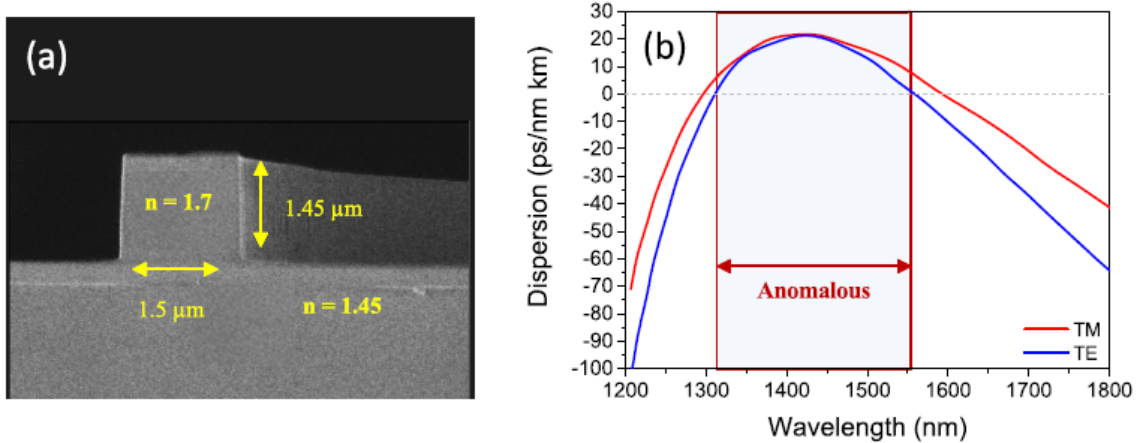

FIG. 1. (a) SEM image of the waveguide prior to over-cladding deposition, (b) estimated dispersion as presented in[16].

The lower nonlinear coefficient, compared to silicon or III-V materials, is compensated for by the significantly lower propagation loss, which leads to a longer effective length. Furthermore, the material bandgap (approx. 6 eV) is much larger than silicon and enables the use of higher pump powers without being limited by the on-set of nonlinear absorption[14,16]. Concerning the impact of Brillouin scattering, the relatively low refractive index difference between core and cladding, is expected to result in poor confinement of the acoustic mode, similar to the silicon-on-insulator platform[19–22]. Additionally, no signs of Brillouin scattering were detected during the measurements reported here, confirming our expectation. Finally, the material is fully CMOS compatible as it can be grown and processed at temperatures below 400° C by using materials already commonly used in CMOS processes.

## III. CW FOUR-WAVE-MIXING CHARACTERIZATION

The nonlinear properties of the spiral waveguide were first characterized with a CW signal by using the setup shown in Fig. 2.

A weak CW signal was coupled into the waveguide together with a strong CW pump (linewidth < 100 kHz). An optical spectrum analyzer (OSA) was used to measure the power of the generated idler at the waveguide output and its value was maximized by aligning the polarization of the two input waves to the TM mode of the waveguide with polarization controllers (PCs).

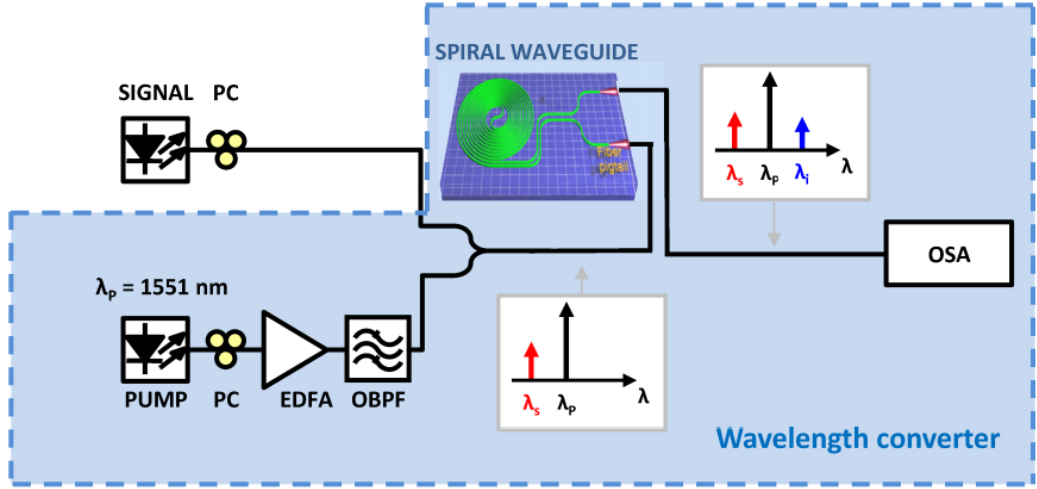

FIG. 2. Experimental setup for the static (CW) FWM characterization of the spiral waveguide.

The conversion efficiency (CE), defined as the power ratio between idler and signal at the waveguide output, is shown in Fig. 3(a) as a function of the input pump power (measured at the input of the fiber pigtail) for a signal-pump spacing of 1 nm and a pump wavelength of 1555 nm. The CE shows the expected quadratic scaling with the pump power. The deviation from linearity above 28-dBm pump power is recoverable and it is believed to be due to heating effects in the glue holding the fiber pigtails in place, since no saturation due to two photon absorption was observed at much higher peak powers at low repetition rates[16]. Fig. 3(b) shows the CE spectrum as a function of the signal wavelength for 30-dBm of pump power and a pump wavelength of 1555 nm. Under such conditions a 3-dB FWM bandwidth of approximately 11 nm is achieved. In order to optimize the CE bandwidth, the pump wavelength needs to be optimized according to the dispersion profile of the waveguide (Fig. 1(b)). Fig. 3(c) addresses this aspect by showing the CE spectra for three different pump wavelengths. Notice that only the half-spectra are shown, i.e. for a signal wavelength shorter than the pump wavelength, as a symmetric shape is expected. By reducing the pump wavelength from 1560 nm down to 1550 nm, the 3-dB FWM half-bandwidth increases from 5.0 nm to 6.3 nm. Additionally, if CE variations of up to 3.5 dB are allowed, by setting the pump wavelength at 1550 nm, an effective half-bandwidth of up to 18 nm could be considered for system applications, covering more than the telecommunication C-band.

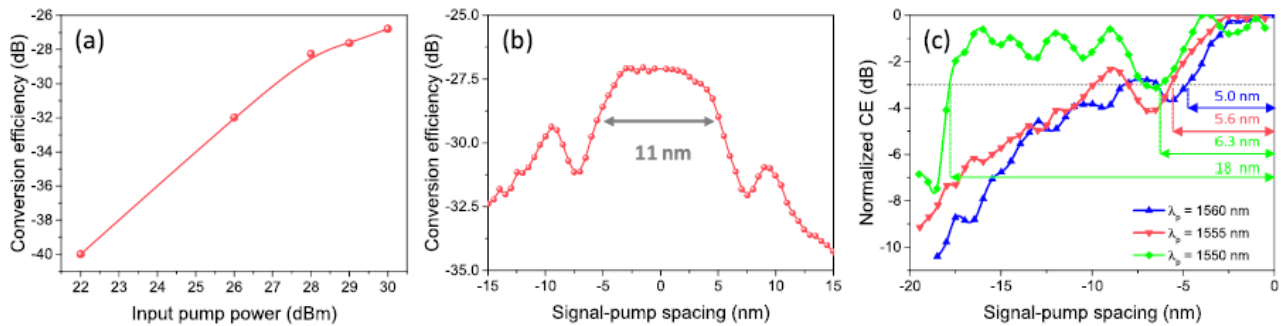

FIG. 3. (a) Conversion efficiency (CE) as a function of the input pump power, (b) full CE versus signal wavelength for a pump power of 30 dBm and a pump wavelength of 1555 nm (partly repro- duced from Ref.[18]) and (c) normalized CE half-spectra for different pump wavelengths and pump power of 30 dBm.

## IV. WAVELENGTH CONVERSION SETUP

The experimental setup for the wavelength conversion of QAM signals is shown in Fig. 4.

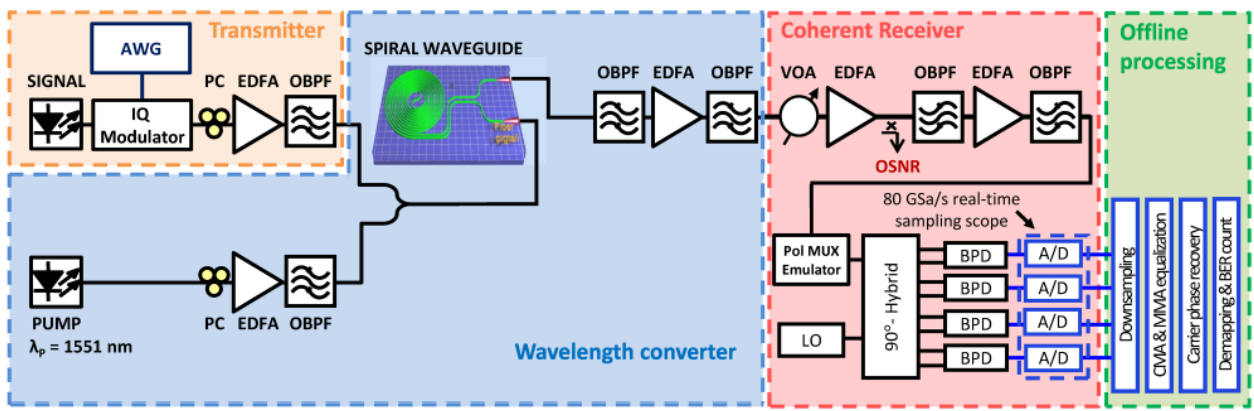

FIG. 4. Experimental setup used for the wavelength conversion of QPSK- and 16-QAM modulated signals.

Both a QPSK and a 16-QAM optical signal were generated by modulating a CW laser using a standard IQ modulator driven by a 64-G sample/s arbitrary waveform generator (AWG) with 20 GHz of analog bandwidth. The pump was provided by another CW laser, set close to 1550 nm to broaden the CE bandwidth as discussed in Section III. However, due to equipment limitations, the pump could only be set to 1551 nm. The pump and signal waves were separately amplified in erbium-doped fiber amplifiers (EDFAs) and the amplified spontaneous emission noise (ASE) was filtered out by optical band-pass filters (OBPFs) with bandwidths of 0.8 nm and 0.5 nm for the signal and pump beams, respectively. The two waves were then coupled into the spiral waveguide with their states of polarization aligned to the TM mode of the waveguide by maximizing the conversion efficiency (CE). At the output of the waveguide the idler wave was selected with a 2-nm wide OBPF, pre-amplified and then filtered again with a 1-nm wide OBPF to fully suppress the remnants of pump and signal waves.

A standard pre-amplified coherent receiver was used to evaluate the quality of the generated idlers which was then compared with the back-to-back signal performance measured directly at the transmitter output. The receiver consisted of a noise loading stage, where a variable optical attenuator (VOA) in front of an EDFA was used to vary the OSNR of the wave under test. The noise-loaded signal (idler) was then pre-amplified and input into a coherent receiver based on a dual-polarization 90° hybrid and four 40-GHz balanced photodiodes (BPDs). A delay-and-add polarization emulator was used to generate a dual-polarization signal compatible with our receiver configuration. A digital sampling oscilloscope (80 Gsamples/s, 33 GHz analog bandwidth) provided the analog-to-digital (A/D) conversion and the acquired waveforms were processed by standard offline digital signal processing (DSP)[23]. The DSP chain consisted of down-sampling, adaptive time-domain equalization using blind radius directed equalizer (RDE), carrier recovery by decision-directed phase lock loop, de-mapping and bit error ratio (BER) counting.

## V.   RESULTS

The results for the 32-GBd QPSK signals are shown in Fig. 5(a) and 5(b). The measurements were performed at moderate pump (22 dBm) and signal (16.5 dBm) powers since the idler OSNR was already sufficient to achieve performance below the HD-FEC threshold. The power levels refer to the input of the fiber pigtail, therefore the actual power in the waveguide is further decreased by splicing loss, coupling loss and mode converter loss.

The optical spectra at the output of the waveguide are shown in Fig. 5(a) as the input signal wavelength is varied between 1549 nm and 1541 nm with the signal-pump spacing at 2 nm, 5 nm and 10 nm. The idler power decreases as the signal is moved further away from the pump with an approximately 3-dB difference between the idler at 1553 nm and the one at 1561 nm, as expected from the static conversion bandwidth measurements of Fig. 3(c).The CE values measured on the spectra at the output of the waveguide are -40.1 dB, -43 dB and -42.7 dB for a signal wavelength of 1549 nm, 1546 nm and 1541 nm, respectively.

Despite the relatively modest CE values, Fig. 5(b) shows that the BER versus OSNR performance of the integrated wavelength converter with the idler BER benchmarked against the back-to-back signal measured directly at the output of the IQ modulator (Dave – is what?? Sentence is only half finished…).  The arrows in the figure point to the maximum idler OSNR achievable for the different signal wavelengths and show a linear decrease in the idler OSNR with the output idler power as the signal-pump spacing is increased.  Nevertheless, all the idlers reach a BER below the HD-FEC threshold with less than 0.3 dB of OSNR penalty over the whole OSNR range available for the measurements.

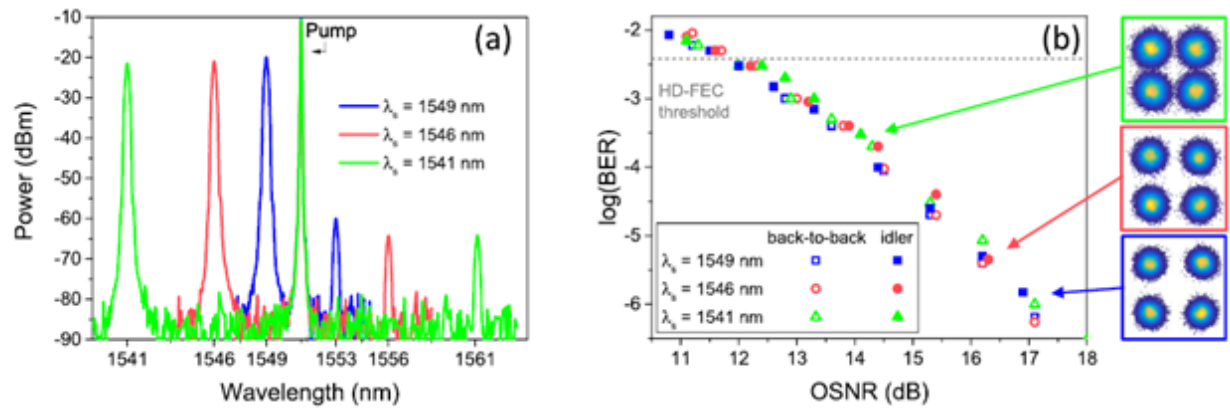

FIG. 5. Wavelength conversion of 32-GBd QPSK: (a) optical spectra at the output of the waveguide and (b) BER performances as a function of the received OSNR. The inset in (b) shows the idler constellation diagrams at the maximum received OSNR for all the three idlers. Arrows relate constellation diagrams and corresponding BER values. This figure is partly reproduced from Ref.[17].

The wavelength converter was then tested using a 10-GBd 16-QAM signal and the results are shown in Fig. 6(a) and 6(b). Given the higher OSNR requirements of 16-QAM, the BER is limited to higher values as compared to the QPSK case.  However, also in this case the idler still achieves a BER below the HD-FEC threshold with an OSNR penalty below 0.3 dB over whole OSNR range and for a signal-pump spacing up to 10 nm, i.e. a wavelength shift of the data signal of up to 20 nm.

Finally, as a benchmark, the wavelength conversion was tested using a straight 3- cm long waveguide. The spectrum at the output of the waveguide for an input signal at 1546 nm and modulated with 32-GBd QPSK is shown in Fig. 7 (a). As expected, the CE is reduced by using a shorter waveguide, however, the lower coupling loss (total insertion loss of approx. 3 dB) compared to the 50-cm waveguide still enables the successful reception of the idler. The BER performance v e r s u s the received OSNR is shown in Fig. 7b, highlighting yet again a very negligible OSNR penalty of 0.2 dB compared to back-to-back at the HD-FEC threshold. In this case though, the idler OSNR is only barely sufficient to achieve a BER below the HD-FEC threshold. In contrast, the wavelength converter based on the longer spiral waveguide provided a safer OSNR margin at the receiver side. Further decreasing the insertion loss of the 50-cm waveguide, particularly the coupling loss which dominates for short lengths, could potentially improve on that margin, as discussed in the following section.

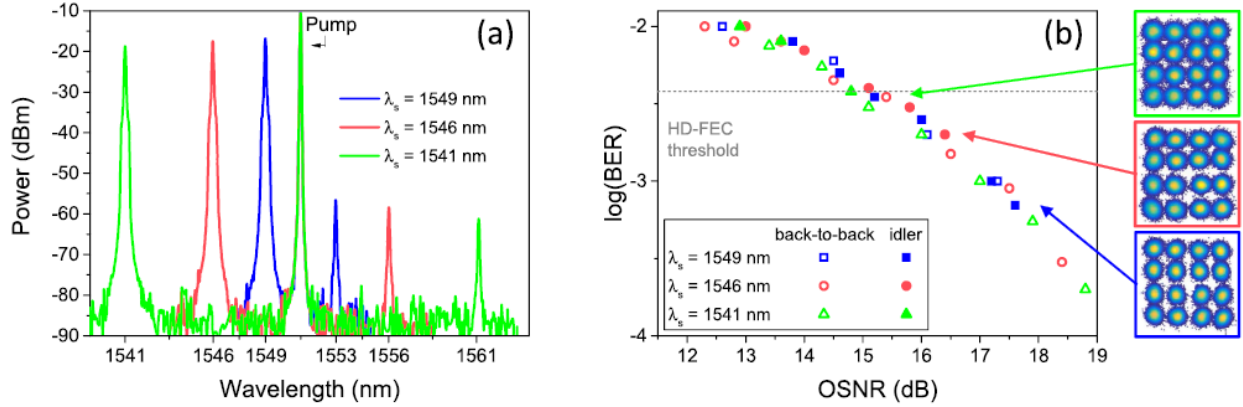

FIG. 6. Wavelength conversion of 10-GBd 16-QAM: (a) optical spectra at the output of the waveguide and (b) BER performances as a function of the received OSNR. The inset in (b) shows the idler constellation diagrams at the maximum received OSNR for all the three idlers. Arrows relate constellation diagrams and corresponding BER values. This figure is partly reproduced from Ref.[17].

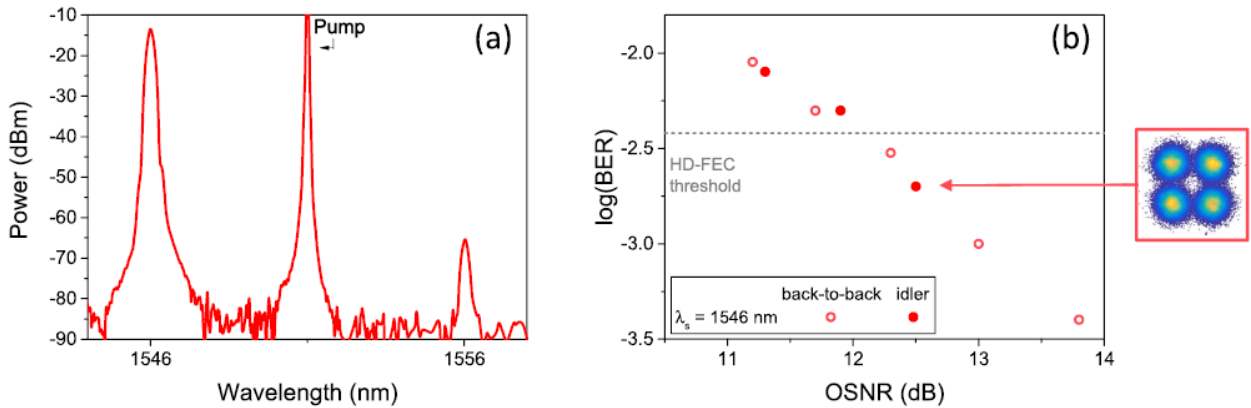

FIG. 7. Wavelength conversion of 32-GBd QPSK signal at 1546 nm in a 3-cm long waveguide: a) optical spectra at the output of the waveguide and (b) BER performances as a function of the received OSNR. The inset in (b) shows the idler constellation diagram at the maximum received OSNR for the considered idler.

## VI. DISCUSSION

The BER results reported in the previous section are summarized in Fig. 8, where the OSNR penalty at the HD-FEC threshold is shown for all the scenarios reported in Section V. The penalty has been calculated by considering the difference in required OSNR between wavelength converted idler and back-to-back signal to reach a BER = HD-FEC threshold.

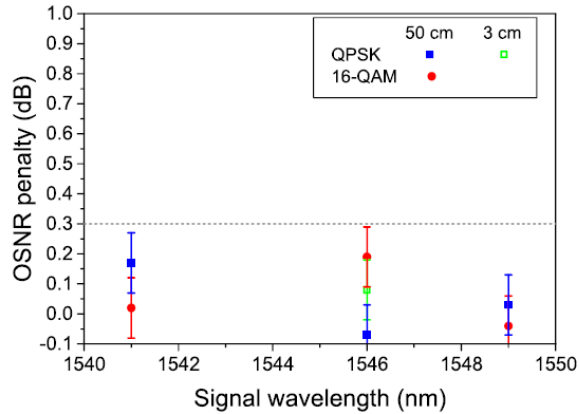

FIG. 8. OSNR penalty as a function of the signal wavelength for the two modulation formats (32- GBd QPSK and 10-GBd 16-QAM) and the two waveguide lengths (50-cm and 3-cm) investigated at the HD-FEC BER threshold. The error bars indicate the uncertainty in OSNR measurement.

As can be seen in Fig. 8, for all the cases considered, 32-GBd QPSK and 10-GBd 16-QAM as well as the two waveguide lengths, the OSNR penalty is kept well below 0.3 dB, a rather negligible value from a systems perspective. Such a penalty may actually be mainly caused by uncertainty in the OSNR measurements, here represented by error bars.

In a wavelength converter, several noise and distortion sources could lead to a non-negligible OSNR penalty after conversion. In particular, the impact of Kerr nonlinearity such as self- and cross-phase modulation can degrade the signal quality through nonlinear distortion, especially when considering wavelength conversion of QAM signal. In the considered waveguide then, the more modest nonlinearity acts as an asset in limiting the nonlinear distortion affecting the signal while still providing a sufficient idler OSNR for reaching a BER below the HD-FEC threshold.

Noise sources within wavelength converters are typically mainly a result of the transfer of phase and intensity noise from the pump to the idler. In this case, however, since the CW pump was of high quality with low intensity noise and a narrow linewidth, these noise transfer effects had little impact, as highlighted by the negligible penalty (Fig. 8). The overall idler power at the output of the waveguide is critical to enable low-noise amplification of the newly generated wave prior to transmission or reception. Having too low a power at the output of the waveguide may irreparably degrade the idler OSNR after amplification. The results of Section V clearly show that the idler power is sufficient to provide enough OSNR to achieve performance below the HD-FEC threshold after amplification and reception. In this regard, the spiral waveguide provides a clear advantage in OSNR margin as compared to the 3-cm waveguide. The waveguide length could even be increased further up to the effective length, i.e. 62 cm considering the propagation loss of 0.07 dB/cm.

We note that, based on the CE versus wavelength shown in Fig. 3(b) and the quadratic scaling shown elsewhere[15], by increasing the pump power up to 30 dBm, we estimate that the maximum idler OSNR available at the receiver could be further increased. That would directly result in a significant OSNR margin that could enable transmission of the idler after the wavelength converter. In our systems investigation, due to the thermal limitation of the glue used for the pigtails and the fact that the available OSNR already allowed performance below the HD- FEC threshold, the pump power was simply kept at 22 dBm. Furthermore, a higher pump power could potentially allow extending the bandwidth over which the available OSNR enables performance better than the HD-FEC threshold. This would, however, result in variations in OSNR and thus the performance over the different wavelength channels, as highlighted by the conversion efficiency bandwidth of Fig. 3(c), and this may result in a slightly increased OSNR penalty if distortion due to cross-phase modulation becomes relevant.

Finally, by decreasing the coupling loss into the waveguide, the idler OSNR is expected to be further increased both by a higher effective pump power coupled into the waveguide and by enabling a higher idler power at the waveguide output. The negligible penalty introduced by the wavelength conversion (< 0.3 dB), together with the sufficiently broad CE bandwidth (20 nm) as well as the potential for improvement by using higher pump powers and achieving lower coupling loss, make this material platform an interesting candidate for implementing all-optical signal processing.

## VII. CONCLUSION

We demonstrate wavelength conversion of 32-GBd QPSK and 10-GBd 16-QAM signals in a 50-cm long, low-loss high-index doped glass waveguide. We achieve an idler BER below the HD-FEC threshold at moderate pump power (22 dBm) for up to 20-nm of wavelength shift between signal and idler, featured by OSNR penalties below 0.3 dB with respect to the original back-to-back signal. Under CW pumping conditions, the maximum CE showed a linear increase with pump power, with a 3-dB CE bandwidth of approx. 11 nm that could be increased up to 36 nm if slightly large CE variations (approx. 3.5 dB) can be tolerated.

## ACKNOWLEDGMENTS

This work was supported by the Center of Excellence Silicon Photonics for Optical Communications (SPOC) funded by the Danish National Research Foundation (ref. DNRF123), by the National Science and Engineering Research Council in Canada, and by the Australian Research Council (ARC) discovery projects program.